\documentclass[aps,pre,twocolumn,floatfix,showpacs,reprint]{revtex4-1}

\usepackage{graphicx}
\usepackage{color}
\usepackage{setspace}
\usepackage{subfigure}
\usepackage{amsmath,times}
\usepackage{pstricks,pst-plot}
\usepackage{upgreek}
\usepackage{hyperref}

\newcommand{\outline}[1]{}

\newcommand{\unit}[1]{\,\mathrm{ #1 }}
\newcommand{\rnd}[1]{\left( #1 \right)}
\newcommand{\of}[1]{\! \left( #1 \right)}

\newcommand{\Pb}{P_{\text{B}}}

\begin{document}

\title{Bubble nucleation in stout beers}

\author{W. T. Lee}
\homepage{www.ul.ie/wlee}
\email{william.lee@ul.ie}
\affiliation{MACSI, Department of Mathematics and Statistics, University of
  Limerick, Limerick, Ireland}

\author{J. S. McKechnie}
\affiliation{MACSI, Department of Mathematics and Statistics, University of
  Limerick, Limerick, Ireland}

\author{M. G. Devereux}
\affiliation{MACSI, Department of Mathematics and Statistics, University of
  Limerick, Limerick, Ireland}

\begin{abstract}
Bubble nucleation in weakly supersaturated solutions of carbon
dioxide---such as champagne, sparkling wines and carbonated beers---is
well understood.  Bubbles grow and detach from nucleation sites: gas
pockets trapped within hollow cellulose fibres. This mechanism appears
not to be active in stout beers that are supersaturated solutions of
nitrogen and carbon dioxide. In their canned forms these beers require
additional technology (widgets) to release the bubbles which will form
the head of the beer. We extend the mathematical model of bubble
nucleation in carbonated liquids to the case of two gasses and show
that this nucleation mechanism is active in stout beers, though
substantially slower than in carbonated beers and confirm this by
observation. A rough calculation suggests that despite the slowness
of the process, applying a coating of hollow porous fibres to the
inside of a can or bottle could be a potential replacement for
widgets.
\end{abstract}
\maketitle

\section{Introduction}

The production of bubbles in weakly supersaturated solutions of carbon
dioxide is of great interest to the beverage industry. Such solutions
include many soft drinks and beers, as well as sparkling wines and
champagne. While it has long been appreciated that spontaneous bubble
formation in these liquids is strongly inhibited and thus that bubble
formation can only occur at certain nucleation
sites~\cite{Walker1981,Jones1999}, it is only comparatively recently
that the nature of these sites has been fully elucidated. In a series
of papers, Liger-Belair and co-workers demonstrated that the most
important nucleation sites are pockets of gas trapped in cellulose
fibres~\cite{Liger2002} (an example of type IV nucleation in the
classification of Jones et al.~\cite{Jones1999}) and developed a
mathematical model of the growth and detachment of these
bubbles~\cite{Liger2005b}, (a complementary model making slighly
different assumptions was developed by Uzel et al.~\cite{Uzel2006}).

While most beers are carbonated, there are advantages to using a
mixture of nitrogen and carbon dioxide in beers, as is done in a
number of stouts. (Hereafter, the term `stout' will be used to refer
to a beer containing a mixture of dissolved nitrogen and carbon
dioxide.)  These advantages include lower acidity in the beer leading
to an improved taste; and smaller bubbles giving a creamy mouthfeel
and a long lasting head~\cite{Bamforth2004,Denny2009}. These beers are
interesting scientifically because they show interesting fluid
dynamical phenomena such as roll waves~\cite{Robinson2008} and sinking
bubbles~\cite{Zhang2008}.  Also of scientific interest is the
technology used to create the head in the canned products.

Pouring a carbonated beer from the can into a glass is enough to
generate the head. This is not the case for stouts. Foaming in canned
stouts is promoted by a widget: a hollow ball containing pressurised
gas. When the can is opened, the widget depressurises by releasing a
gas jet into the beer. The jet breaks up into tiny bubbles which are
carried throughout the liquid by the turbulent flow generated by the
gas jet and by pouring the beer into a glass. Dissolved gasses diffuse
from the liquid into the bubbles which rise to the surface of the beer
to form the head.

In this paper we extend the mathematical model of bubble formation in
carbonated liquids developed by Liger-Belair et al.~\cite{Liger2005b}
to the case of two dissolved gasses and use it to investigate two
questions:
\begin{itemize}
\item{Why do stout beers require widgets? Is the bubbling mechanism
  described by Liger-Belair et al.\ completely inactive in stout beers
  or merely very slow?}
\item{Could an alternative to the widget be developed by coating part
  of the inside of the can by hollow fibres?}
\end{itemize}

\begin{table}
\caption{\label{parameters} Values of parameters used in this
  work. }
\begin{center}
\begin{tabular}{c r@{}l c}
\hline
Parameter & \multicolumn{2}{c}{Value} & Reference\\
\hline
\hline
$r$                 &  $6.00\times 10^{-6}$&$\unit{m}$ & \cite{Liger2005b}\\
$\lambda$           & $14.00\times 10^{-6}$&$\unit{m}$ & \cite{Liger2005b}\\
$\gamma$            & $47.00\times 10^{-3}$&$\unit{N}\unit{m}^{-1}$ 
                                                       & \cite{Bamforth2004} \\
$D_1$ & $1.40\times 10^{-9}$&$ \unit{m}^2 \unit{s}^{-1}$\\ 
$D_2$ & $2.00\times 10^{-9}$&$ \unit{m}^2 \unit{s}^{-1}$\\ 
$H_1$ & $3.4\times 10^{-4}$&$\unit{mol}\unit{m}^{-1}\unit{N}^{-1}$ 
                                                       & \\
$H_2$ & $6.1\times 10^{-6}$&$\unit{mol}\unit{m}^{-1}\unit{N}^{-1}$
                                                       & \\
$T$   & $293$&$\unit{K}$ \\
$P_0$ & $1.00\times 10^{5}$&$ \unit{Pa}$\\ 
$P_1$ & $0.80\times 10^{5}$&$ \unit{Pa}$ & \cite{ESGI70}\\ 
$P_2$ & $3.00\times 10^{5}$&$ \unit{Pa}$ & \cite{ESGI70}\\
\hline
\end{tabular}
\end{center}
\end{table}

\section{Mathematical Model}

In this section we develop a mathematical model of the rate of growth
of a gas pocket in a cellulose fibre for the case in which there are
two dissolved gasses: nitrogen and carbon dioxide. Once a gas pocket
reaches a critical size (when it reaches an opening of the fibre) it
rapidly forms a bubble outside the fibre, leaving behind the original
gas pocket. Since bubble formation and detachment is much faster than
the growth of the gas pocket, the rate at which bubbles are nucleated
can be deduced from the rate of growth of the gas
pocket~\cite{Liger2005b}.

\begin{figure}
\begin{center}
\includegraphics[]{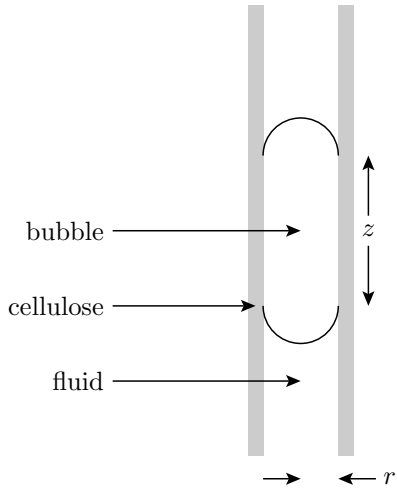}
\end{center}
\caption{\label{geometry} Geometry of a gas pocket trapped in a
  cellulose fibre. }
\end{figure}

The geometry of a gas pocket in a cellulose fibre is shown in
Fig.~\ref{geometry}.  Dissolved gasses in the fluid
diffuse into the bubble through the walls of the cellulose fibre and
through the spherical caps at the ends of the gas pocket. The rate at
which this occurs is determined by the surface area, the diffusion
constant and a diffusion length scale. The diffusion constants used to
calculate the fluxes of carbon dioxide and nitrogen through the
spherical caps are the diffusion constants in free fluid: $D_1$ and
$D_2$. The relevant diffusion constants for flow through the cellulose
walls are $D_{1\perp}$ and $D_{2\perp}$. NMR measurements show that
for carbon dioxide $D_{1\perp}\approx 0.2D_1$~\cite{Liger2004}. We
assume the same relationship holds between $D_{2\perp}$ and $D_2$. The
diffusion lengthscale $\lambda$ was measured experimentally for carbon
dioxide~\cite{Liger2005b}, again we assume that this value is also
valid for nitrogen diffusion.

In this model the rate of change of the numbers of carbon dioxide
($N_1^*$) and nitrogen ($N_2^*$) molecules in the gas pocket are given
by
\begin{align} 
\dfrac{\text{d} N_1^*}{\text{d} t^*} &= 4\pi r^2 D_1 \dfrac{\Delta c_1}{\lambda}
                           +2\pi rz D_{1\perp} \dfrac{\Delta c_1}{\lambda}, \\
\dfrac{\text{d} N_2^*}{\text{d} t^*} &= 4\pi r^2 D_2 \dfrac{\Delta c_2}{\lambda}
                           +2\pi rz D_{2\perp} \dfrac{\Delta c_2}{\lambda},
\end{align}
where asterisks indicate dimensional variables that will be
non-dimensionalised later.

Using Henry's law, Laplace's law and the ideal gas equation:
\begin{align}
\Delta c_1 &= H_1\rnd{P_1 -\dfrac{\Pb N_1^*}{N_1^*+N_2^*}},\\
\Delta c_2 &= H_2\rnd{P_2 -\dfrac{\Pb N_2^*}{N_1^*+N_2^*}}.\\
\Pb &= P_0+\dfrac{2\gamma}{r}, \\
z &= \dfrac{\rnd{N_1^*+N_2^*} RT}{\pi r^2\Pb},
\end{align}
where $P_1$ is the partial pressure of dissolved carbon dioxide, $P_2$
is partial pressure of dissolved nitrogen, $\Pb$ is the pressure in
the gas pocket given by the Laplace law, $P_0$ is atmospheric pressure
and $\gamma$ surface tension.

These equations can be non-dimensionalised by using the scales
\begin{gather}
N_{\text{scale}}=\dfrac{2D_2}{D_{2\perp}}  \dfrac{\Pb \pi r^3}{RT}
                \approx 3.22\times 10^{-13}\unit{mol},  \\
t_{\text{scale}}=\dfrac{r\Pb\lambda}{2D_{2\perp} H_2 P_2 RT}
                \approx 2.73\unit{s},
\end{gather}
to introduce dimensionless variables $N_1$, $N_2$ and $t$ (by
assumption $D_2/D_{2\perp}=D_1/D_{1\perp}$). The dimensionless
equations are
\begin{align}
\label{dimensionless_one}
\epsilon \dfrac{\text{d} N_1}{\text{d} t} 
  &=\rnd{1+N_1+N_2}\rnd{1-\dfrac{\alpha_1N_1}{N_1+N_2}}, \\
\label{dimensionless_two}
\dfrac{\text{d} N_2}{\text{d} t}
  &=\rnd{1+N_1+N_2}\rnd{1-\dfrac{\alpha_2N_2}{N_1+N_2}}.
\end{align}
Using values typical of stouts, given in Table~\ref{parameters}, the
dimensionless parameters are
\begin{gather}
\epsilon = \dfrac{D_2H_2P_2}{D_1H_1P_1}\approx 0.096, \\
\alpha_1 = \dfrac{\Pb}{P_1}\approx 1.45,\\
\alpha_2 = \dfrac{\Pb}{P_2}\approx 0.39.
\end{gather}

\section{Asymptotic Solution}

Equations~\ref{dimensionless_one} and~\ref{dimensionless_two} cannot
be solved directly. They can, however, be solved in two asymptotic
limits: $\epsilon\ll 1$ and $N_1+N_2\gg 1$. The former limit does not
produce particularly accurate results but the analysis of this limit
helps us to interpret the results from taking the second asymptotic
limit. The results from taking the second asymptotic limit are more
accurate but harder to understand intuitively.

\subsection{First asymptotic limit: $\boldsymbol{\epsilon\ll 1}$}

Taking the limit in which the small parameter $\epsilon\approx 0.1$ is
zero, equation~\ref{dimensionless_one} becomes an algebraic equation
\begin{equation}
\label{algebraic}
0=1-\dfrac{\alpha_1N_1}{N_1+N_2},
\end{equation}
which can be substituted into
equation~\ref{dimensionless_two}
\begin{equation}
\dfrac{\text{d} N_2}{\text{d} t}
  =  \dfrac{\alpha_1+\alpha_2-\alpha_1\alpha_2}{\alpha_1-1}N_2
    +\dfrac{\alpha_1+\alpha_2-\alpha_1\alpha_2}{\alpha_1}.
\end{equation}
This equation is solved by
\begin{equation}
N_2= A\exp\of{\dfrac{t}{\tau}}
     -\dfrac{\alpha_1+\alpha_2-\alpha_1\alpha_2}{\alpha_1},
\end{equation}
where $A$ is a constant of integration and $\tau$ a dimensionless time
constant describing the timescale of growth of the gas pocket in this
approximation
\begin{gather}
 \tau = \dfrac{\alpha_1-1}{\alpha_1+\alpha_2-\alpha_1\alpha_2}
     \approx 0.35,\\
\tau t_{\text{scale}} =0.954\unit{s}.
\end{gather}

Physically this approximation corresponds to assuming that diffusion
of carbon dioxide is infinitely fast, and thus the partial pressure of
carbon dioxide in the gas pocket is always equal to the partial pressure
of carbon dioxide in solution. Obviously this approximation is only
valid if the partial pressure of carbon dioxide is less than the
gas pocket pressure, otherwise equation~\ref{algebraic} has no physical
solutions. Note that this approximation will underestimate $\tau$
since it assumes carbon dioxide diffusion is infinitely fast.

\subsection{Second asymptotic limit: $\boldsymbol{N_1+N_2\gg 1}$}

\begin{table}
\caption{\label{eigenvectors} Numerical values of the parameters in
  equations~\ref{eva} and~\ref{evb}.}
\begin{center}
\begin{tabular}{c r@{.}l}
\hline
Parameter   &   \multicolumn{2}{c}{Value}  \\
\hline\hline
$a_{11}$ &   0&989 \\
$a_{12}$ &   0&836 \\
$a_{21}$ & --0&145 \\
$a_{22}$ &   0&548 \\
$\tau_1$ &   0&161 \\
$\tau_2$ &   0&468 \\
$\tau_1 t_{\text{scale}}$  &  0&439$\,$s \\
$\tau_2 t_{\text{scale}}$  &  1&278$\,$s \\
\hline
\end{tabular}
\end{center}
\end{table}

In the limit $N_1+N_2\gg 1$ equations~\ref{dimensionless_one}
and~\ref{dimensionless_two} become 
\begin{align}
\dfrac{\text{d} N_1}{\text{d} t}&=-\dfrac{\alpha_1-1}{\epsilon}N_1+\dfrac{N_2}{\epsilon},\\
\dfrac{\text{d} N_2}{\text{d} t}&= N_1+\rnd{1-\alpha_2}N_2.
\end{align}
These equations have two independent solutions
\begin{equation}
\label{eva}
N_1=Aa_{11}\exp\of{-\dfrac{t}{\tau_1}},
\quad
N_2=Aa_{21}\exp\of{-\dfrac{t}{\tau_1}},
\end{equation}
and 
\begin{equation}
\label{evb}
N_1=Ba_{12}\exp\of{\dfrac{t}{\tau_2}},
\quad
N_2=Ba_{22}\exp\of{\dfrac{t}{\tau_2}},
\end{equation}
where $A$ and $B$ can be chosen independently to satisfy initial
conditions. The numerical values of the other parameters are given in
Table~\ref{eigenvectors}. 

The analysis of the $\epsilon\ll 1$ case allows us to interpret these
two solutions. The first solution, equation~\ref{eva}, decays
exponentially with a small timescale. This corresponds to the rapid
establishment of the (dynamic) equilibrium concentrations (or partial
pressures) of CO$_2$ and N$_2$ within the gas pocket (assumed
instantaneous in the previous analysis). The second solution,
equation~\ref{evb}, shows exponential growth with a longer
timescale. This describes the steady state growth of the gas pocket at
a fixed concentration ratio of CO$_2$ to N$_2$. The timescale of this
process describes the timescale of bubble production. This analysis
produces a longer estimate of that timescale than the previous
analysis. This is because, previously, diffusion of CO$_2$ was assumed
to be instantaneous. As the numerical results described in the next
section show, the $\epsilon\ll 1$ limit underestimates the correct
timescale, while the $N_1+N_2\gg 1$ analysis gives a good estimate.

\section{Numerical Solution}

\begin{figure}
\begin{center}
\includegraphics{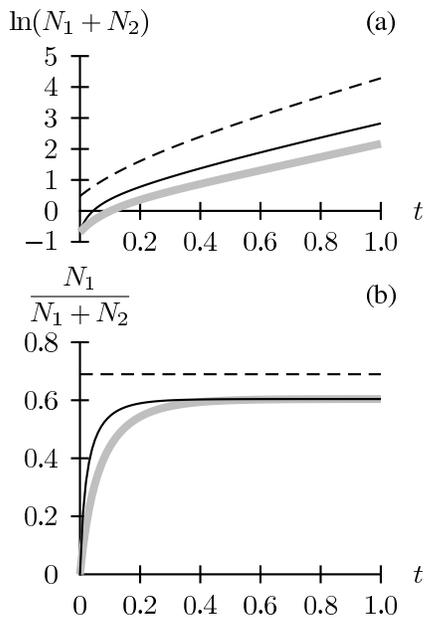}
\end{center}
\caption{\label{graphs} Results of the numerical solution of
  equations~\ref{dimensionless_one} and~\ref{dimensionless_two}. The
  black line shows the numerical solution, the grey line shows the
  $N_1+N_2\gg 1$ limit, and the dashed black line shows the
  $\epsilon\ll 1$ limit. (a) Rate of growth of the gas pocket. (b)
  Evolution of the concentration of CO$_2$ in the gas pocket.}
\end{figure}

\begin{figure*}
\begin{center}
\includegraphics{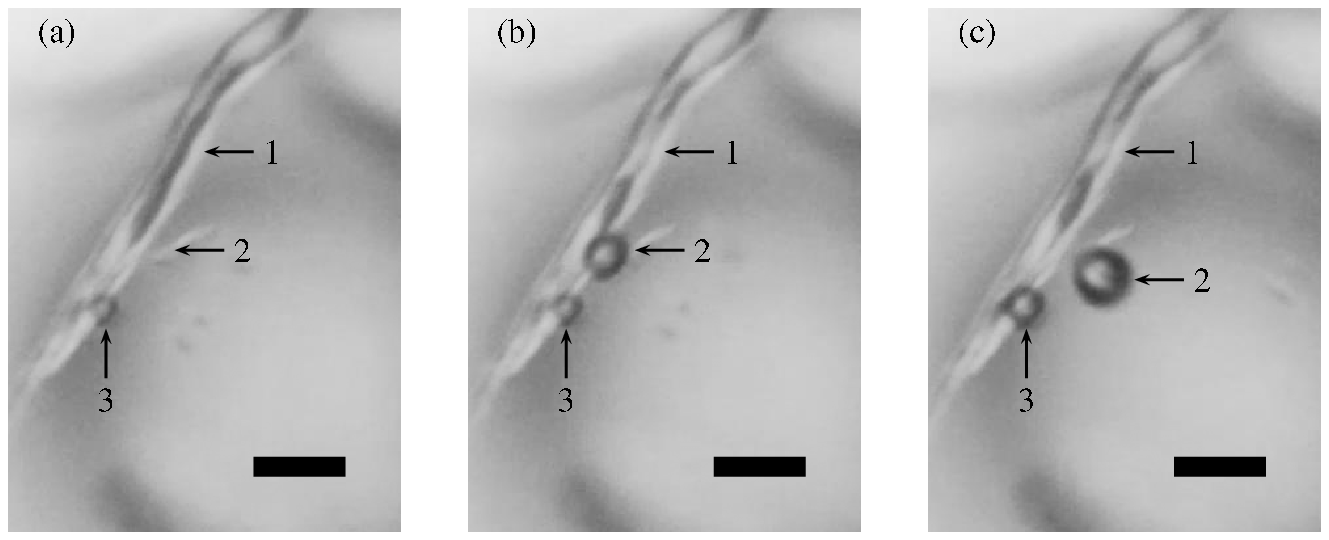}
\end{center}
\caption{\label{fibre}Bubble nucleation in stout from a
  cellulose fibre taken from a coffee filter. The scalebar is
  $50\,\upmu\text{m}$ in each figure. The arrows show: 1~a gas pocket
  in the fibre which nucleates a bubble; 2~the bubble fed by gas from
  pocket~1; 3~a bubble nucleated on the outside of the
  fibre. (a)~The air pocket (1) has reached maximum size. (b)~The air
  pocket (1) has created bubble (2). (c)~Bubble (2) has visibly
  detached from the fibre, air pocket (1) is starting to refill with
  gas. (a), (b) and (c) are frames from a movie. (b) is $80\unit{ms}$
  after (a), (c) is $520\unit{ms}$ after (a).}
\end{figure*}

A full solution of the dimensionless equations can be obtained by
numerical integration.  A fourth order Runge-Kutta scheme with a
timestep of $10^{-3}$ was used to solve
equations~\ref{dimensionless_one} and~\ref{dimensionless_two} with
initial conditions $N_1=0$, $N_2=0.5$. The differential equations were
solved over the interval $0<t<10$. The result for $N=N_1+N_2$ for
$5<t<10$ were fitted to an exponential curve giving a dimensionless
bubble growth timescale of $\tau=0.47$ corresponding to a dimensional
timescale of $\tau t_{\text{scale}}=1.28\unit{s}$, in agreement with
that predicted from the analysis of the $N_1+N_2\gg 1$ case. This can
be compared with the value for carbonated liquids at the same total
pressure: $0.079\unit{s}$. Fig.~\ref{graphs} shows the results of
the numerical simulations over the interval $0<t<1$.

In conclusion, these analytic and numerical results suggest that the
mechanism of bubble formation described by Liger-Belair et al.\ is
potentially active in stout beers but acts much more slowly than in
carbonated drinks.

\section{Experimental Confirmation}

\begin{figure}
\begin{center}
\includegraphics{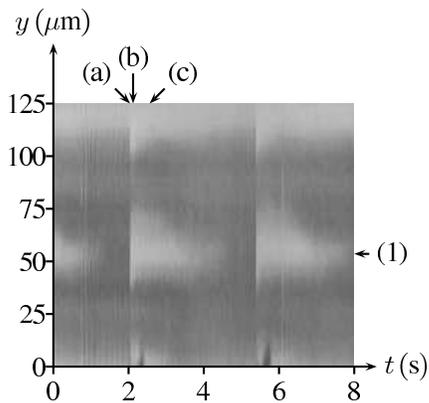}
\end{center}
\caption{\label{gas_pocket}Growth of the gas pocket. This figure shows
  the growth of the gas pocket within the cellulose fibre shown in
  Fig.~\ref{fibre}. As in Fig.~\ref{fibre} dark colours correspond
  to gas and light colours to liquid. The columns of pixels from the
  frames corresponding to parts (a), (b) and (c) of Fig.~\ref{fibre}
  are indicated. The location of the gas pocket indicated by a (1) in
  Fig.~\ref{fibre} is shown. }
\end{figure}

To confirm experimentally that cellulose fibres could nucleate bubbles
in stout beer, we observed a canned draught stout in contact with
cellulose fibres (taken from a coffee filter). Before opening the can,
we made a small hole in the can to slowly degass the widget. This
prevented it from foaming, which would have removed the dissolved
gasses from solution.  Using a microscope we observed that bubbles
were indeed nucleated from the cellulose fibres but at a relatively
slow rate.  Fig.~\ref{fibre} shows bubbles nucleated by a gas
pocket: the three parts of Fig.~\ref{fibre} are frames taken from a
movie.

Fig.~\ref{gas_pocket} shows the growth of the gas pocket shown in
Fig.~\ref{fibre}. The figure was constructed from the same movie of
the bubbling process used for Fig.~\ref{fibre}. Two hundred frames,
corresponding to $8\unit{s}$ were extracted from the movie and rotated
so that the fibre shown in Fig.~\ref{fibre} was vertical. From each
frame the same column of pixels, passing through the centre of the
fibre, was extracted and those columns placed side by side to construct
a new figure: Fig.~\ref{gas_pocket}. This figure shows the evolution
of the gas pocket: its slow growth (as predicted by the model) and
then its rapid loss of gas to form an external bubble (as assumed by
the model).

\section{Widget Alternatives}

The model developed above allows us to investigate the feasibility of
an alternative foaming strategy for stout beers in cans and bottles in
which a coating of hydrophobic fibres on the inside of the can is used
to promote foaming.  A typical pouring time for a stout beer is
$30\unit{s}$. In this time about $10^8$ postcritical nuclei must be
released. A single fibre produces one bubble every
$1.28\unit{s}$. Therefore about $4.3\times 10^6$ fibres are needed. If
each fibre occupies a surface of area $\lambda^2$ then the total area
that must be occupied by fibres is $8.3\times 10^{-4}\unit{m}^2$
equivalent to a square with edge length $2.9\unit{cm}$. This indicates
that such an approach may be possible.

\section{Conclusions}

A model of bubble formation in carbonated liquids has been extended to
the case of liquids containing both dissolved nitrogen and carbon
dioxide. Taking values typical of stout beers shows that bubble
formation by this mechanism does occur but at a substantially slower
rate. This is consistent with the observation that widgets are needed
to promote foaming in canned stouts. The possibility of replacing
widgets with an array of hollow fibre nucleation sites was
investigated and shown to be potentially feasible.

\begin{acknowledgments}
We acknowledge support of the Mathematics Applications Consortium for
Science and Industry (\url{www.macsi.ul.ie}) funded by the Science
Foundation Ireland Mathematics Initiative Grant 06/MI/005. MD
acknowledges funding from the Irish Research Council for Science,
Engineering and Technology (IRCSET).
\end{acknowledgments}

\end{document}